\documentclass[prl,aps,floatfix,twocolumn,superscriptaddress]{revtex4}
\usepackage{graphicx}
\usepackage{epsfig}
\begin{document}

\title{Thermodynamics of carrier-mediated magnetism in semiconductors}

\author{A.~G.~Petukhov}
\affiliation{Department of Physics, South Dakota School of Mines and
Technology, Rapid City, SD 57701}
\author{Igor \v{Z}uti\'{c}}
\affiliation{Department of Physics, State University of New York at Buffalo, 
Buffalo, NY 14260}
\author{Steven C. Erwin}
\affiliation{Center for Computational Materials Science, Naval Research 
Laboratory, Washington, DC 20375}

\begin{abstract}
We propose a model of carrier-mediated ferromagnetism in
semiconductors that accounts for the temperature dependence of the
carriers.  The model permits analysis of the thermodynamic stability
of competing magnetic states, opening the door to the construction of
magnetic phase diagrams.  As an example we analyze the stability of a
possible {\em reentrant} ferromagnetic semiconductor, in which 
increasing temperature leads to an increased carrier density, such that
the enhanced exchange coupling between magnetic impurities results in
the onset of ferromagnetism as temperature is raised.
\end{abstract}
\maketitle 

Ferromagnetic semiconductors (FS) show important and potentially
useful differences from their metallic counterparts.  For example, if
the magnetism in a magnetically doped semiconductor is mediated by carriers, then
changes in the carrier density induced by light or applied bias may
significantly alter the the exchange interaction between the carriers
and magnetic impurities.  When this effect is sufficient to turn the
ferromagnetism on or off, there arise intriguing possibilities for
light- or bias-controlled
ferromagnetism~\cite{Koshihara1997:PRL,Ohno2000:N,Park2002:S,Zutic2004:RMP}
not possible in conventional ferromagnetic metals.

Another approach to controlling ferromagnetism in FS materials is to
exploit the strong temperature dependence of carrier density
that is the hallmark of semiconductors.  Despite this dependence, most
theoretical analysis of FS has assumed a metallic-like
picture~\cite{Dietl2001:PRB,Jungwirth2006:RMP} in which carrier density is treated
as independent of temperature.  Consequently, a number of experimental
features observed in certain FS materials remain incompletely
explained.  Examples include the metal-insulator
transition~\cite{VanEsch1997:PRB} in Mn-doped GaAs; the role of the
impurity band in oxide FS materials~\cite{Berciu2001:PRL,Coey2005:NM};
and the resistivity peak in Mn-doped GaAs observed near the Curie
temperature, $T_C$, which is usually attributed to temperature-dependent
scattering~\cite{Ohno1998:S} just as in metals~\cite{Fisher1968:PRL}.
In each of these cases,  correct treatment of the temperature dependence of
the carrier density could substantively change our understanding of
the observed phenomena.

In this paper we develop a theoretical model of ferromagnetism in
semiconductors that includes the temperature dependence of the
carriers.  By providing a way to analyze the thermodynamic stability
of different competing magnetic states, this model makes possible the
self-consistent calculation of the magnetic phase diagram of a FS.
Here we use the model to calculate the temperature-dependent
magnetization of a simple generic FS. In contrast to the standard
monotonic decay of the magnetization with increasing temperature found
in metals, we demonstrate the possibility of stable ``reentrant''
ferromagnetism in semiconductors: as the temperature is increased the
resulting higher density of thermally excited carriers can enhance the
exchange coupling between magnetic impurities---and thereby {\em
  increase} the magnetization over some range of temperatures.  Of
course, whether such a possibility can be realized even in principle
depends on the thermodynamic stability of the reentrant phase relative
to other magnetic states. To properly analyze this competition
requires a theoretical framework for computing the free energy of each
possible magnetic state. Our model provides this framework.

\begin{figure}[tbph]
\centerline{\psfig{file=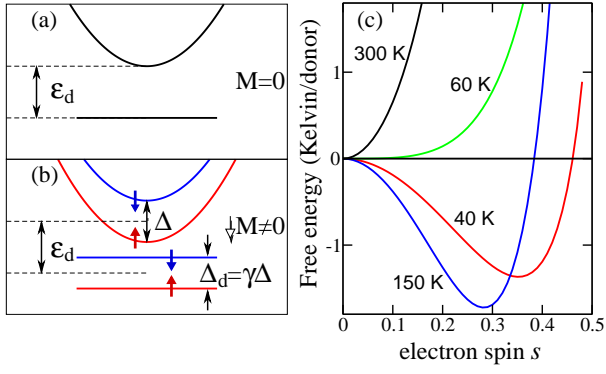,width=1\linewidth,angle=0}}
\caption{Schematic energy-band structure,
  and temperature evolution of free energy. (a) In the absence of
  magnetization ($M=0)$ both the conduction band and the donor level
  (at $-\varepsilon_d$ relative to the band edge) are unsplit.  (b)
  The onset of magnetic order ($M \neq 0$) leads to spin splitting
  $\Delta$ of the conduction band and $\Delta_d$ of the donor level .
  Thick arrows represent two spin projections.  (c) Free energy $\cal
  F$ versus average electron spin $s$.  Materials parameters
  were chosen based on Gd-doped EuO~\cite{Mauger1986:PR}:
  $\Gamma_{ex}=40$ eV$\cdot$\AA$^3$, $|\varepsilon_d|=20$ meV,
  $N_d=3.5 \times 10^{19}$ cm$^{-3}$, $m^*=2$, lattice constant
  $a_0=5.15$ \AA, $N_i=4/a_0^3$, $\gamma=1$, $a_B=8$ \AA, and $J=7/2$.
  Behavior at $40$ and $150$ K reveals ferromagnetic states.  }
\label{sketch}
\end{figure}

We begin by considering a FS doped with donors of density $N_d$ and
with magnetic impurities of density $N_i$. For simplicity we assume
that no
acceptors are present, that electrons are the only carriers, and that
the magnetic impurities are electrically neutral; the more general
case is straightforward.  The interaction between electron spins $\vec
s_i$ and localized impurity spins $\vec J_j$ is taken to be
\begin{equation}
H_{ex}=-\Gamma_{ex}\sum_{i,j} \delta(\vec r_i -\vec R_j)
\vec s_i \cdot \vec J_j,
\label{eq:kondo}
\end{equation}
where $\Gamma_{ex}$ is the exchange coupling and $\vec r_i$ $(\vec
R_j)$ is the position of the carrier (impurity).  We assume a
nondegenerate semiconductor in which the conduction band is
separated from the donor level (or impurity band) by
$\varepsilon_d$ in the absence of magnetic order, as in
Fig.~\ref{sketch}(a). When
magnetic order is present the conduction band and donor level
experience spin splittings of $\Delta$~\cite{Dietl1983:PRB} and
$\Delta_d$~\cite{Haas1968:PR}, respectively, as in
Fig.\ \ref{sketch}(b). For simplicity we assume that the ratio
$\Delta_d/\Delta\equiv\gamma$ has a fixed, material-specific value.

These spin splittings can arise from either an applied magnetic field
or from the carrier-impurity interaction, and in general will modify
the temperature-dependence of the electron
density~\cite{Zutic2006:PRL}.  We can obtain a
expression for the density $n$ of conduction electrons as
follows.   Electroneutrality requires that $n+N_d^0=N_d$,
where $N_d^0$ is the density of neutral (non-ionized) donors.  If we
take the effective density of states in the conduction band as
$N_c=(1/4)(2 m^*k_BT/\pi\hbar^2)^{3/2}$~\cite{Ashcroft:1976}, then the
electroneutrality condition can be satisfied by introducing the
chemical potential $\mu$ that satisfies
$N_c\exp(\mu/k_BT)\cosh(\Delta/2k_BT)= N_d/\left[1+2\exp\left[(\mu+
    \varepsilon_d)/k_BT\right]\cosh(\Delta_d/2k_BT)\right]$.  This is
equivalent to a quadratic equation in the electron density, and has
the solution
\begin{equation}
n(s,T)=\frac{1}{2} N_d \; \kappa(s,T)\left[\sqrt{1+4/\kappa(s,T)}-1\right].
\label{eq:nust}
\end{equation}
Here we have expressed the dependence of $n$ on the splitting $\Delta$
equivalently as a dependence on the average spin of the conduction
electrons, $s=(1/2)\tanh(\Delta/2k_BT)$. We have also 
defined the auxiliary function
$\kappa(s,T)=(N_c/N_d)\exp(-\varepsilon_d/k_BT)
(1-4s^2)^{(\gamma-1)/2}/[(1+2s)^\gamma+(1-2s)^\gamma]$.
In addition, for later use it is here convenient to 
introduce the {\em relative} density of conduction electrons
$\nu(s,T)=n(s,T)/N_d$, so that $0 < \nu(s,T) < 1$.

In the magnetically ordered state, both the conduction electrons and
the electrons bound to donors experience the field from the ordered
impurity spins.  The impurity spins in turn experience both a
long-ranged uniform field from the delocalized conduction electrons
and a short-ranged field from the bound donor electrons. This
situation can be described using two coupled order parameters: the
average spin $s$ of the conduction electrons as defined above;
and the average spin $m=M/N_ig_i \mu_B$ of the impurities
(here $M$ is the total magnetization due to the impurities and $g_i$
is the electron $g$-factor).

Using these two order parameters, we can write the free-energy density
of the system in the form
\begin{equation}
{\cal F} = {\cal F}_i+{\cal F}_e-\Gamma_{ex}N_i N_d
\left[\nu(s,T) s  + \nu_d(s,T)s_d \right]m.
\label{eq:free_two}
\end{equation}
Here ${\cal F}_i={\cal F}_i(m)$ is the contribution from the entropy
of the impurity
spins, and ${\cal F}_e={\cal F}_e(s)$ is the contribution from the
entropy of the conduction electrons.  The former can be obtained by expressing the
free energy of the impurity spins as a function of an external
magnetic field and then applying a Legendre
transformation~\cite{Kubo:1960}.  This gives
\begin{equation}
{\cal F}_i(m)=N_ik_BT\left[2J\alpha(m)
-\ln\frac{\sinh\left
[{\textstyle (2J+1)}\alpha(m)\right]}
{\sinh\left[\alpha(m)\right]}\right],
\label{eq:free_local}
\end{equation}
where $\alpha(m)=B_J^{-1}(m/J)/2J$, and $B_J^{-1}(x)$ denotes the
inverse of the Brillouin function~\cite{Ashcroft:1976}.

The electron contribution ${\cal F}_e(s)$ 
can be derived using a similar approach. We first define the
density-weighted average spin of conduction and donor electrons,
\begin{equation}
\tau(s,T)= \nu(s,T)s+[1-\nu(s,T)]s_d,
\label{eq:tau}
\end{equation}
where the average spin of donor electrons is given by
$s_d=(1/2)[(1+2s)^\gamma-(1-2s)^\gamma]/[(1+2s)^\gamma+(1-2s)^\gamma]$.
We next invert the function $\tau(s,T)$ by solving Eq.\ (\ref{eq:tau})
for $s=s(\tau,T)$.  This function can then be used to obtain the free
energy of the conduction and donor electrons:
\begin{equation}
{\cal F}_e(s)=k_BT N_d\int_0^{\tau(s,T)}\ln\frac{1+2s(\tau^\prime,T)}
{1-2s(\tau^\prime,T)}
d\tau^\prime.
\label{eq:free_electron}
\end{equation}

The third term in Eq.\ (\ref{eq:free_two}) is the mean-field
approximation for the internal energy described by the exchange
Hamiltonian of Eq.\ (\ref{eq:kondo}).  This term represents the
coupling of the order parameters $s$ and $m$. The expression in square
brackets makes explicit the separate contributions from conduction
electrons and donor electrons.  Since the conduction electrons are
delocalized, they mediate a long-ranged interaction between impurity
spins.  In contrast, the interaction between donor electrons and
impurity spins is short-ranged, and is controlled by a contact term
given by the value of the donor wavefunction at the impurity site.  We
turn now to evaluating this interaction by deriving an explicit
expression for the term $\nu_d(s,T)$ appearing in
Eq.\ (\ref{eq:free_two}).

We use a ``two-color'' percolation model to represent the short-ranged
interaction between the randomly distributed donor electrons and
impurity spins~\cite{Ioselevich1995:PRL,Petukhov2002:PRL}.  This model
is a generalization of a more standard percolation approach originally
proposed for dilute ferromagnets~\cite{Shender1975:PL} and recently
applied to magnetic semiconductors~\cite{Kaminski2002:PRL}.  Within
the two-color model, an interaction is counted for each pair of sites whose spatial
separation is less than $R_c=[B_c/(4\pi/3)]^{1/3}(N_i
{N}_d^0)^{-1/6}$, where $B_c\approx 2.7$ is the average coordination
number \cite{Ioselevich1995:PRL}.
The density of such pairs within the infinite percolation network is
$N_{cl}B_c/2$, where $N_{cl}=2/(4\pi R_c/3)^3$ is the average density
of donor- and impurity-sites belonging to the network.

Within the
mean-field approximation, the contribution of these pairs to the
internal energy can be shown to be $U_{cl}=-\Gamma_{ex}|\psi
(0)|^2\exp\left(-2R_c/a_B\right)(1/2)N_{cl} B_c m s_d$, where
$\psi(0)$ is the value of the donor wavefunction at the
origin ($|\psi(0)|^2=1/\pi a_B^3$ for hydrogenic donors).  Comparing
this result to Eq.\ (\ref{eq:free_two}) implies that
\begin{equation}
\nu_d = {|\psi(0)|^2}{(N_i N_d)^{-1/2}}\left(1-\nu\right)^{1/2}
\exp\left(-2R_c/a_B\right).
\label{eq:newnu}
\end{equation}

\begin{figure}[t]
\centerline{\psfig{file=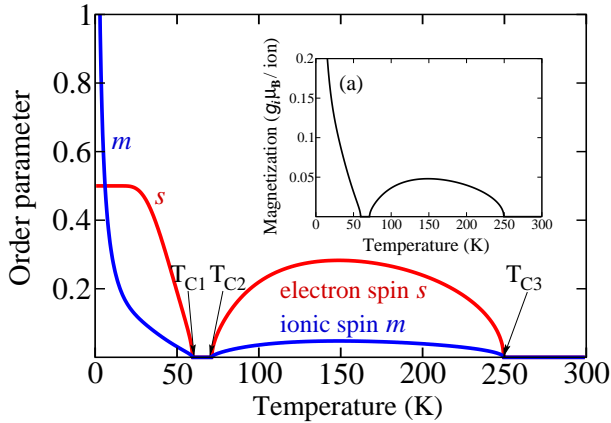,width=1\linewidth,angle=0}}
\caption{Temperature dependence of order parameters $s$ and $m$, 
  showing reentrant behavior at $T=T_{C2}$ as temperature is raised.
  Curie temperatures $T_{C1,2,3}$, at which $m$ and $s$ both vanish,
  define phase boundaries between two ferromagnetic and two
  paramagnetic states.  (a) Inset: Temperature dependence of the
  combined average spin from all electrons.  Materials parameters are
  the same as in Fig.\ \ref{sketch}.  }
\label{OP1}
\end{figure}

Having now obtained explicit expressions for all the terms in
Eq.\ (\ref{eq:free_two}), we can directly obtain the temperature
dependence of the order parameters $s$ and $m$ by minimizing the
free energy $\cal F$.  This results in
\begin{equation}
s=\frac{1}{2}\tanh\left[\frac{\Gamma_{ex}N_im}{2k_BT}
\frac{\partial \left[\tilde\nu(s,T)s\right] }{\partial \tau(s,T)}
\right], 
\label{eq:s_mft}
\end{equation}
\begin{equation}
m=JB_J\left[\frac{J\Gamma_{ex}N_d \tilde\nu(s,T)s}{k_BT}\right],
\label{eq:m_mft}
\end{equation}
where $\tilde\nu(s,T)=\nu(s,T)+\nu_d(s,T)s_d/s$. 
By expanding Eqs.~(\ref{eq:s_mft}) and (\ref{eq:m_mft}) for small 
$s$ and $m$ we obtain an implicit expression for the critical temperature:
\begin{equation}
T_C= T_C^0 \; \tilde\nu(0,T_C)/ 
\left[\gamma + (1-\gamma){\nu(0,T_C)}\right]^{1/2},
\label{eq:TCgamma}
\end{equation}
where $T_C^0=(\Gamma_{ex}/k_B)[N_i N_d J(J+1)/12]^{1/2}$ is the
Curie temperature in the limit of completely ionized
donors~\cite{Coey2005:NM}. 

We now turn to exploring the predictions of our model for a
realistic example.  We choose materials parameters approximately
corresponding to Gd-doped EuO~\cite{Mauger1986:PR}, a magnetic material known
to exhibit strong temperature dependence of the carrier density.
Fig.\ \ref{sketch}(c) shows the resulting free energy $\cal F$ as a
function of conduction-electron spin $s$, at the temperatures 40, 60,
150, and 300 K.  A ferromagnetic state (at 150 K) appears at {\em
  higher} temperature than a paramagnetic state (at 60 K), and is thus
a reentrant ferromagnetic state.  This phenomenon is a direct consequence of
the increased number of thermally excited carriers, which sufficiently increase the
exchange coupling between magnetic impurities to overcome the
additional entropic cost of a magnetically ordered state.

Figure \ \ref{OP1} shows the temperature dependence of $s(T)$ and $m(T)$
for the same materials parameters.  For
this example there are three different solutions of
Eq.\ (\ref{eq:TCgamma}) for $T_C$, as shown. The reentrant ferromagnetic state
exists in the temperature range $T_{C1} \le T \le T_{C2}$.
The inset shows the temperature dependence of the combined average spin from
all electrons (conduction, donor, and impurity), revealing behavior
very different from the conventional monotonic decay~\cite{Ashcroft:1976}.

In the standard theoretical description of dilute magnetic
semiconductors the ferromagnetism is mediated entirely by ``free''
carriers (usually holes in the valence
band)~\cite{Dietl2001:PRB,Jungwirth2006:RMP}.  An important check of
our theory is that it reproduces earlier results obtained in this
limit. We thus consider the case with all donors ionized, $n=N_d$.  In
this regime the entropy contribution ${\cal F}_e(s)$ must be replaced
with its degenerate counterpart for the internal energy, ${\cal F}_e^{\rm deg}(s)=(3n
E_F/10)\left[(1+2s)^{5/3}+(1-2s)^{5/3}\right]$, where
$E_F=(\hbar^2/2m^*)(3\pi^2n)^{2/3}$ is the Fermi energy.  Substituting
this expression into Eq.\ (\ref{eq:free_two}) and expanding ${\cal F}$
in $s$ and $m$ near the Curie temperature, we obtain
$T_C=(\Gamma_{ex}^2/k_B)N_i N_0 J(J+1)/12$, where $N_0=3n/2E_F$ is the
electronic density of states at $E_F$. This is the standard expression
for $T_C$ given, for example, as Eq.\ (7) of Ref.~\cite{Dietl2001:PRB}.

The possibility of reentrant ferromagnetism in semiconductors was first
discussed over forty years ago~\cite{Karpenko1964:FTT}, and again
recently~ \cite{Calderon2006:P}. In neither case was the thermodynamic
stability of the magnetism discussed. Moreover, the role of the donor
electrons was not included, an omission that can lead to
contradictions.  For example, in Ref.\ \cite{Calderon2006:P} it was
assumed that the conduction-electron density has no explicit
dependence on the magnetization.  This assumption is justified only if
the spin-splitting of the donor level ($\Delta_d$) and conduction band
($\Delta$) are equal, i.e. $\gamma=1$. However, the self-consistency
equation given in Ref.\ \cite{Calderon2006:P} for the magnetization of
the impurity spins is only correct for the case $\gamma=0$.  This can
be seen by substituting our Eq.~(\ref{eq:s_mft}) into
Eq.~(\ref{eq:m_mft}) and comparing to the corresponding equation in
Ref.\ \cite{Calderon2006:P}.  This internal inconsistency has
substantial consequences. Specifically, for the case $\gamma=1$,
expanding the self-consistency equation for small magnetization in the
vicinity of the critical temperature yields $T_C= T_C^0\;\nu(0,T_C)$.
But for the case $\gamma=0$, a similar expansion yields the
qualitatively different dependence $T_C= T_C^0 \sqrt{\nu(0,T_C)}.$

\begin{figure}[tbph]
\centerline{\psfig{file=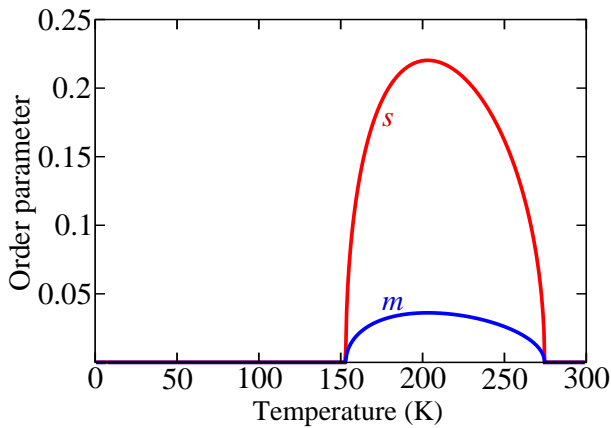,width=1\linewidth,angle=0}}
\caption{Temperature dependence of order parameters $s$ and $m$
  when percolation is omitted (see text).  Materials parameters are
  the same as in Fig.\ \ref{sketch}, except for $m^*$=2.5.}
\label{OP2}
\end{figure}

In its present form, our model does not include several physical
effects which could alter the magnetic ordering at low temperature.
(1) Hopping and Coulomb fluctuations will broaden the donor levels
into an impurity band of localized states~\cite{Shklovskii:1984}.  (2)
The neutral donors may form bound magnetic polarons~\cite{Nagaev:1983}
by aligning nearby impurity spins, thereby increasing the donor
ionization energy $\varepsilon_d$. (3) There could be a substantial
direct antiferromagnetic interaction between nearest-neighbor impurity
spins.  This would have two consequences: ${\cal F}$ in
Eq.\ (\ref{eq:free_two}) would acquire a contribution $\propto m^2$,
while the argument in the mean-field expression for $m$ in
Eq.\ (\ref{eq:m_mft}) would be reduced by a term $\propto m$.

All three of these effects would suppress ferromagnetism at low
temperatures.  While the situation is fairly complex, we can describe
it qualitatively by omitting the percolation term in Eq.\ 
(\ref{eq:free_two}), i.e.~by setting $\nu_d(s,T)=0$ or, equivalently,
by taking $a_B\rightarrow 0$ in Eq.\ (\ref{eq:newnu}).  Figure \ 
\ref{OP2} shows the resulting behavior for $s(T)$ and $m(T)$.  The
striking feature is that the ferromagnetism is absent at both low and
high temperatures, corresponding to antiferromagnetic and paramagnetic
states, respectively. A similar interpretation of reentrant
ferromagnetism was proposed in Ref.\ \cite{Karpenko1964:FTT} to explain experimental data on
(Li,Mn)Se \cite{Pickart1961:PR}.
Furthermore, both of the scenarios for reentrant ferromagnetism shown in 
Figs.~\ref{OP1} and \ref{OP2} 
are consistent with the recent experiments in
(In,Mn)Se~\cite{Slynko2005:PRB}. With the change in Mn-concentration
there is a change in the number of peaks in the temperature dependence
of dynamic magnetic susceptibility supporting the existence of either
two or three distinct critical temperatures (recall Figs.~\ref{OP1} and \ref{OP2}).
While, for simplicity,  we have focused on the parameters for 
Eu-based semiconductors, there is a need to explore other FS for a possible
reentrant ferromagnetism.

Predictions of reentrant ferromagnetism could be directly tested in
transport experiments. For example, in conventional (non-reentrant)
spin light-emitting diodes~\cite{Zutic2004:RMP} the emitted light is
circularly polarized if the injected carriers are spin-polarized.  As
the temperature is raised, there is a monotonic decrease in the degree
of polarization of the electroluminescence, due to the decreased
polarization of the carriers.  By using a reentrant FS as the spin
injector, a nonmonotonic temperature dependence of the
electroluminescence, similar to that for $s(T)$ in Figs.~\ref{OP1} or
\ref{OP2}, would be expected. Reentrant behavior could also be
detected electrically, without a spin-LED, by demonstrating 
non-monotonic temperature dependence of the magnetoresistance in a
semiconductor heterojunction~\cite{Zutic2006:PRL,Chen2006:PRB} that
includes a reentrant FS.

This work was supported by the US ONR, NSF-ECCS CAREER, CNMS at ORNL,
and the CCR at SUNY Buffalo.

\end{document}